\documentclass[conference]{IEEEtran}

\usepackage{amsmath,amssymb,amsfonts}
\usepackage{graphicx}
\graphicspath{{./}{../}{../../}}   % ./figs first, else the shared paper/figs/
\usepackage{booktabs}
\usepackage{multirow}
\usepackage{tikz}
\usetikzlibrary{arrows.meta,positioning,shapes.geometric}
\usepackage{xcolor}
\usepackage[bookmarks=false,hidelinks]{hyperref}

\usepackage{cite}

\begin{document}

\title{Semantic Communication for Distributed Spectrum Monitoring over Unreliable Links}

\author{\IEEEauthorblockN{Samer Lahoud}
\IEEEauthorblockA{\textit{Faculty of Computer Science} \\
\textit{Dalhousie University}\\
Halifax, NS, B3H 4R2 Canada \\
sml@dal.ca}
\and
\IEEEauthorblockN{Kinda Khawam}
\IEEEauthorblockA{\textit{ROCS, LISN} \\
\textit{Paris-Saclay University}\\
91190 Gif-sur-Yvette, France \\
kinda.khawam@universite-paris-saclay.fr}
}

\maketitle

\begin{abstract}
	Distributed spectrum monitoring relies on spatially separated receivers to infer the state of the radio environment. A central task is to determine how many emitters are active and where they are located. Sensors typically report over shared unreliable links, where contention, fading, or energy limits erase whole reports, so forwarding raw I/Q samples from all receivers to a fusion node becomes impractical. This paper formulates distributed emitter counting and localization as a semantic communication problem over unreliable reporting links. Each receiver maps its local time-frequency observation to a compact latent representation and transmits it. The fusion node pairs each surviving latent with the corresponding receiver position and operates on the resulting unordered set, which keeps the fusion rule compatible with variable membership and receiver-count changes. The encoder and decoder are trained end-to-end with report erasure, and optionally finite-rate quantization, inside the task objective. Experiments on synthetic multi-emitter scenes show that channel-aware training improves robustness under erasure, that set fusion is more robust than fixed-order concatenation, and that one trained model operates across receiver counts without retraining. The rate study shows that a few bits per latent component suffice in the tested setting, and that the rate knee remains roughly stable across erasure levels. These results support the feasibility of low-rate semantic reporting for distributed spectrum monitoring over unreliable links.
\end{abstract}


\section{Introduction}
\label{sec:intro}

Semantic and task-oriented communication shift the design objective from reproducing data to delivering the information needed for a downstream task
\cite{gunduz2023beyond,strinati2021beyond,shao2022ib}. Distributed spectrum monitoring is a natural setting for this shift. A monitoring system must infer the state of the radio environment, including how many emitters are active and where they are located. This information supports dynamic spectrum access, interference management, anomaly detection, and security in beyond-5G and 6G networks. In many deployments, monitoring is performed by low-cost spatially distributed receivers. Forwarding raw in-phase/quadrature (I/Q) streams from all receivers to a fusion node then costs bandwidth, energy, and access-channel resources, so each receiver should transmit a compact task-relevant report instead.

The communication link that carries these reports is also part of the problem. Monitoring receivers often report over shared, contention-based, or grant-free links rather than over dedicated scheduled channels. In such settings, a collision, a deep fade, or an energy limitation erases a whole report
\cite{abramson1970aloha,shahab2020grantfree}, so the impairment is packet-level as well as bit-level. A model trained on complete report sets can therefore degrade sharply when the fusion node receives only a subset of the receivers. The reporting set also changes across deployments, and from one observation window to the next. A practical semantic communication scheme for spectrum monitoring must handle both whole-report erasure and variable receiver membership.

This paper studies emitter counting and localization as a concrete distributed spectrum-monitoring task. 
%These outputs are deliberately chosen because they depend on spatial diversity. The spectral and waveform parameters of the emitters affect the received signals and may create spectral overlap, but they are treated here as nuisance variables rather than outputs. 
The objective is to estimate the emitter count and locations from the reports that survive the unreliable links. Our approach is to align the learning architecture with the reporting channel. Each receiver applies a shared encoder to its local time-frequency observation and sends the resulting compact latent. The fusion node pairs each survivor with its known receiver position and processes the reports as an unordered set. This makes the decoder invariant to the order of the received reports and compatible with different report memberships. We also make the training channel-aware by inserting whole-report erasure, and optionally finite-rate quantization, inside the end-to-end task objective.

The main contributions are as follows:
\begin{itemize}
	\item We cast distributed emitter counting and localization over unreliable reporting links as a channel-aware semantic communication problem. The learned representation is optimized for task distortion under whole-report erasure and finite-rate constraints.
	
	\item We introduce a permutation-invariant fusion architecture that operates on the surviving set of position-tagged receiver reports. This makes the decoder tolerant of variable report membership and enables cross-count deployment without retraining.
	
	\item We evaluate the effects of channel-aware training, fusion architecture, receiver-count variation, and latent quantization. The results show improved robustness under erasure, stronger performance than fixed-order concatenation, and accurate operation with only a few bits per latent component.
	
\end{itemize}

The remainder of the paper is organized as follows. Section~\ref{sec:related} reviews related work. Section~\ref{sec:system} presents the system model and channel-aware objective. Section~\ref{sec:method} describes the encoder, set-fusion decoder, loss function, and training procedure. Section~\ref{sec:experiments} presents the experimental results, and Section~\ref{sec:conclusion} concludes the paper.

\section{Related Work}
\label{sec:related}

\paragraph{Task-oriented and semantic communication}
Task-oriented and semantic communication replace bit-level reconstruction by task utility as the design objective
\cite{gunduz2023beyond,strinati2021beyond,xie2021dl}. The information bottleneck principle provides a useful way to trade representation rate against task relevance
\cite{tishby2000ib}. This idea has been applied to edge inference with a single device
\cite{shao2022ib} and to cooperative inference across multiple devices through the distributed information bottleneck
\cite{shao2022dib}. Related rate-distortion formulations also make the goal-oriented trade-off explicit
\cite{stavrou2022rd}. Our work follows this task-oriented view and focuses on a distributed monitoring setting in which the reporting links are unreliable and the set of received reports is variable. The channel is therefore part of the learning problem, rather than a rate constraint imposed after the representation has been designed.

\paragraph{Deep joint source-channel coding}
Deep joint source-channel coding maps a source directly to channel symbols through a learned autoencoder and can provide graceful degradation as channel quality changes
\cite{bourtsoulatze2019djscc}. Nonlinear-transform variants further connect source-channel coding with semantic objectives
\cite{dai2022ntscc}. These works are closest to learned end-to-end communication. The setting considered here differs in one respect. Multiple receivers send separate task-oriented reports over unreliable reporting links, where the main impairment is the erasure of whole reports. We therefore keep a digital reporting interface and model report loss directly.

\paragraph{Learned RF sensing and distributed compression} Deep learning has been used for spectrum sensing, wireless signal classification, and transmitter localization from distributed measurements
\cite{rajendran2018dl,zhan2021deepmtl}. Most of these systems assume that the measurements or learned features reach the fusion point reliably. A closely related recent work studies task-oriented compression for multi-emitter localization and characterization with spectral overlap
\cite{distcomp2026}. We build on this direction and move from a clean link, fixed-array setting to unreliable reporting links and variable receiver membership. Classical cooperative spectrum sensing also studies the effect of quantization, bandwidth constraints, and reporting errors on fusion
\cite{akyildiz2011coop,sun2007coop}. Compressive sensing reduces the acquisition cost for wideband monitoring
\cite{tian2007compressed}, while in-network aggregation reduces the amount of data sent by sensor networks
\cite{madden2002tag}. In contrast, our approach learns the semantic representation under the same erasure and rate constraints that affect deployment.

\paragraph{Permutation-invariant set models}
The proposed fusion rule draws on permutation-invariant neural architectures such as Deep Sets
\cite{zaheer2017deepsets} and attention-based set pooling
\cite{lee2019settransformer}. In our setting, the reporting channel dictates this architectural choice. The fusion node receives whichever subset of reports survives contention, fading, or energy limitations. Treating the inputs as a set therefore makes the decoder invariant to report ordering, tolerant of missing reports, and compatible with receiver counts different from the one used during training.

% ==== Core technical sections ====
% Section III: System Model  (LOCALIZATION paper)
% Requires: \usepackage{amsmath}
\section{System Model} 
\label{sec:system}

\subsection{Monitoring Scenario and Task}
\label{sec:scenario}

A two-dimensional monitoring region $\mathcal{A}\subset\mathbb{R}^2$ is observed by $M$ sensing nodes. Node $i$ has known position $r_i\in\mathcal{A}$, $i\in\{1,\dots,M\}$, which reaches the decoder as side information. In a deployment this is slow-varying metadata, registered once when a node joins rather than repeated in every observation window. Positions are redrawn for every scene, so the model handles an arbitrary receiver layout.

The region contains an unknown number $K$ of simultaneously active emitters, with $1\le K\le K_{\max}$. Emitter $k$ has location $p_k\in\mathcal{A}$ and physical signal parameters
\begin{equation}
	\xi_k=(\nu_k,B_k,P_k,g_k),
\end{equation}
where $\nu_k$ is the center-frequency offset, $B_k$ is the occupied bandwidth, $P_k$ is the transmit power, and $g_k$ denotes the waveform family. These parameters shape the received signals and often create spectral overlap between emitters. They remain nuisance variables in this work. The monitoring output is the emitter set
\begin{equation}
	\Theta=\{p_k\}_{k=1}^{K},
\end{equation}
together with its cardinality $K$.

This choice focuses the study on counting and localization, which are central tasks in distributed spectrum monitoring. Both draw on spatial diversity, so every lost report removes one spatial view of the same emitter scene. The task therefore probes reporting-link reliability directly. Since $\Theta$ is an unordered set with unknown cardinality, a permutation-invariant set distortion $D_{\mathrm{set}}(\Theta,\hat{\Theta})$ scores the estimate $\hat{\Theta}$. This distortion charges both cardinality errors and localization errors after the best assignment between predicted and true emitters, as described in Section~\ref{sec:loss}.

\begin{figure}[t]
  \centering
  \begin{tikzpicture}[scale=0.74, >=Latex, font=\footnotesize,
    rx/.style={regular polygon, regular polygon sides=3, draw, fill=blue!25,
               inner sep=1.3pt},
    em/.style={star, star points=5, star point ratio=2.3, draw, fill=orange!85,
               inner sep=1.2pt}]
    \draw[thick] (0,0) rectangle (8,8);
    % propagation links: every emitter to every receiver
    \begin{scope}[gray!55, dashed, line width=0.3pt]
      \foreach \e in {(3,4.2),(5.6,2.7),(4.4,6.3)}{
        \foreach \r in {(0.6,1.1),(1.3,7.0),(7.1,1.5),(6.6,6.1)}{
          \draw \e -- \r;}}
    \end{scope}
    % receivers
    \node[rx,label={[font=\scriptsize]below:$r_1$}] at (0.6,1.1) {};
    \node[rx,label={[font=\scriptsize]above:$r_2$}] at (1.3,7.0) {};
    \node[rx,label={[font=\scriptsize]below:$r_3$}] at (7.1,1.5) {};
    \node[rx,label={[font=\scriptsize]above:$r_4$}] at (6.6,6.1) {};
    % emitters
    \node[em,label={[font=\scriptsize]right:$p_1$}] at (3,4.2) {};
    \node[em,label={[font=\scriptsize]right:$p_2$}] at (5.6,2.7) {};
    \node[em,label={[font=\scriptsize]right:$p_3$}] at (4.4,6.3) {};
    % legend
    \node[rx] at (0.3,-0.7) {};
    \node[anchor=west,font=\scriptsize] at (0.55,-0.7) {receiver $r_i$ (position known)};
    \node[em] at (0.3,-1.2) {};
    \node[anchor=west,font=\scriptsize] at (0.55,-1.2) {emitter $p_k$};
  \end{tikzpicture}
\caption{Monitoring geometry. A top-down $1000\times1000$~m region $\mathcal{A}$ contains several active emitters at unknown locations, shown as stars, and $M$ receivers with known but scene-dependent positions, shown as triangles. Dashed lines denote emitter-to-receiver propagation. Each receiver observes a different noisy superposition of the active emitters, producing a local window $x_i$. The task is to estimate the number of emitters and their locations.}
  \label{fig:scenario}
\end{figure}
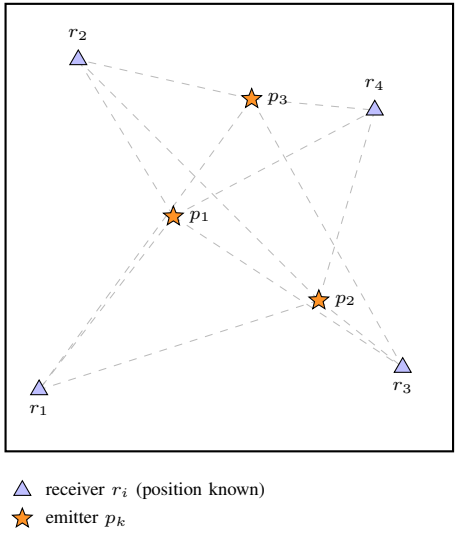

Fig.~\ref{fig:scenario} illustrates the monitoring geometry. Receivers have known but scene-dependent positions, while the emitter locations and signal parameters remain unknown. Each receiver observes a different noisy superposition of the active emitters and reports a compact representation to the fusion node.

\subsection{Distributed Semantic Encoding}
\label{sec:encoding}

The fusion node recovers just enough information to estimate the emitter count and locations, which frees the reports from reconstructing the raw receiver observations. This makes distributed spectrum monitoring a task-oriented, or semantic, communication problem
\cite{gunduz2023beyond,strinati2021beyond,shao2022ib}.

Each receiver observes a window of $N$ complex baseband samples $x_i\in\mathbb{C}^N$. It first computes a time-frequency feature
\begin{equation}
	X_i=\mathcal{T}(x_i),
\end{equation}
and then maps this feature to a compact latent representation
\begin{equation}
	z_i=\mathrm{Enc}_{\phi}(X_i)\in\mathbb{R}^{d}.
\end{equation}
The same encoder $\mathrm{Enc}_{\phi}$ is shared by all receivers. This choice reflects the fact that receiver positions vary across scenes and deployments. Every receiver plays the same role, so a shared extractor keeps the model applicable to any node, while specialization by receiver index would tie it to one geometry.

Only the latent $z_i$ crosses the reporting link. At the fusion node it is paired with the known position of its originating receiver, forming the report token
\begin{equation}
	u_i=(z_i,r_i).
\end{equation}
The position tag is essential because the same latent has a different geometric meaning depending on where the observation was taken. The fusion node therefore assembles task-oriented reports rather than raw signals. The encoder and decoder are trained end-to-end using the set distortion $D_{\mathrm{set}}$, so the latent $z_i$ is optimized for emitter counting and localization rather than for reconstructing $x_i$.

\begin{figure*}[t]
	\centering
	\resizebox{\textwidth}{!}{%
		\begin{tikzpicture}[
			font=\footnotesize, >=Latex,
			blk/.style={draw, rounded corners, align=center, minimum height=8mm,
				minimum width=15mm, inner sep=3pt},
			sig/.style={font=\scriptsize, inner sep=1pt}]
			
			% emitter scene
			\node[blk, minimum height=20mm] (scene) at (0,0.05) {Emitter scene\\ $\Theta=\{p_k\}$};
			
			% per-receiver encoder branches
			\node[blk] (e1) at (3.5,2.1) {Rx $1$: $\mathcal{T}+\mathrm{Enc}_{\phi}$};
			\node[blk] (e2) at (3.5,0.5) {Rx $2$: $\mathcal{T}+\mathrm{Enc}_{\phi}$};
			\node      (ed) at (3.5,-0.7) {$\vdots$};
			\node[blk] (eM) at (3.5,-2.0) {Rx $M$: $\mathcal{T}+\mathrm{Enc}_{\phi}$};
			
			% tokens
			\node[blk] (t1) at (6.6,2.1) {$z_1$};
			\node[blk] (t2) at (6.6,0.5) {$z_2$};
			\node[blk] (tM) at (6.6,-2.0) {$z_M$};
			
			% erasure link
			\node[blk, minimum height=45mm, fill=red!5] (up) at (8.9,0.05) {Erasure\\ reporting link\\ $\epsilon$};
			
			% set-fusion decoder
			\node[blk, minimum height=22mm, align=center] (dec) at (12.3,0.05)
			{Permutation-invariant\\ set fusion\\ $\mathrm{Dec}_{\psi}(\{(z_i,r_i)\}_{i\in\mathcal{S}})$};
			
			% outputs
			\node[blk, minimum height=20mm, align=left] (out) at (15.5,0.05)
			{$\hat\Theta$:\\ count $\hat K$\\ locations\\ $\{\hat p_k\}$};
			
			% scene -> encoders
			\draw[->] (scene.east) -- (e1.west) node[sig, pos=0.6, above]{$x_1$};
			\draw[->] (scene.east) -- (e2.west) node[sig, pos=0.6, above]{$x_2$};
			\draw[->] (scene.east) -- (eM.west) node[sig, pos=0.6, below]{$x_M$};
			
			% encoders -> tokens
			\draw[->] (e1) -- (t1);
			\draw[->] (e2) -- (t2);
			\draw[->] (eM) -- (tM);
			
			% tokens -> link
			\draw[->] (t1) -- (up.west|-t1);
			\draw[->] (t2) -- (up.west|-t2);
			\draw[->, gray, dashed] (tM) -- node[red, pos=0.6]{\large$\times$} (up.west|-tM);
			
			% link -> decoder -> outputs
			\draw[->] (up) -- (dec) node[sig, midway, above]{$\{z_i\}_{i\in\mathcal{S}}$};
			\draw[->] (dec) -- (out) node[sig, midway, above]{$\hat\Theta$};
		\end{tikzpicture}%
	}
	\caption{Distributed semantic reporting and set fusion. The emitter scene $\Theta$ is observed by $M$ receivers at scene-dependent positions. Each receiver applies the same encoder to its time-frequency feature and transmits the resulting compact latent $z_i$ over an unreliable reporting link, which may erase whole reports with probability $\epsilon$; one erased report is shown by $\times$. At the fusion node, each surviving latent is paired with its known receiver position $r_i$ to form $u_i=(z_i,r_i)$, and a permutation-invariant decoder maps the surviving set to the emitter count and locations.}
	\label{fig:pipeline}
\end{figure*}
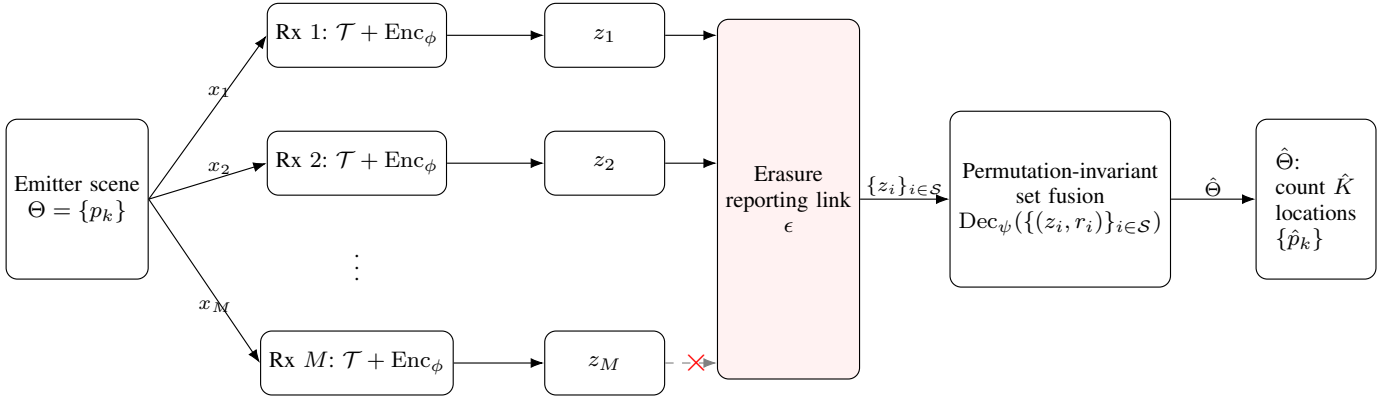

Fig.~\ref{fig:pipeline} summarizes the reporting pipeline. Each receiver applies the shared encoder locally and sends its latent over an unreliable reporting link. The next subsection describes how the fusion node combines the reports that survive this link.

\subsection{Permutation-Invariant Fusion over Surviving Reports}
\label{sec:fusion}

The reporting link described in Section~\ref{sec:uplink} delivers a random subset $\mathcal{S}\subseteq\{1,\dots,M\}$. The fusion node therefore receives a variable-size collection whose membership the channel determines, in place of a fixed-length ordered vector. We model this collection as a set and write
\begin{equation}
	\hat\Theta=\mathrm{Dec}_\psi\big(\{u_i:i\in\mathcal{S}\}\big).
	\label{eq:setfusion}
\end{equation}
The decoder is permutation-invariant, so its output stays the same under any reordering of the surviving reports
\cite{zaheer2017deepsets,lee2019settransformer}.

This is the central architectural choice. The unordered set sits on the reporting side as well as the output side, so a lost report simply leaves the input set instead of occupying a placeholder in a fixed vector. This matches the behavior of contention-based and unreliable reporting links, where the channel determines which receivers are observed by the fusion node.

The same decoder is therefore well defined for any surviving subset $\mathcal{S}$. It can process different report memberships from one window to the next, and it can also be evaluated with receiver counts different from the one used during training. A fixed-order concatenation decoder ties its input dimension to a prescribed receiver count and index ordering, so it applies at its own training count alone. Section~\ref{sec:decoder} gives the implementation of $\mathrm{Dec}_\psi$ using a shared per-report embedding followed by masked attention pooling.

\subsection{Finite-Rate Semantic Reporting}
\label{sec:source}

The latent vector $z_i$ is the semantic message sent by receiver $i$. To model a finite-rate reporting link, each component of $z_i$ is quantized to $b$ bits. We use per-dimension affine min--max calibration: each latent component is mapped to the unit interval, uniformly quantized, and then mapped back to the latent scale. As $b\to\infty$, this operation approaches the full-precision latent.

The per-receiver semantic rate and the offered reporting load per observation window are
\begin{equation}
	R_i=bd,\qquad
	R_{\mathrm{tot}}=\sum_{i=1}^{M} R_i=Mbd
	\quad\text{[bits/window]}.
\end{equation}
The rate is therefore controlled by the latent dimension $d$ and the bit depth $b$. Receiver positions arrive as side information, so $R_i$ counts the semantic payload alone. In the experiments that isolate erasure and fusion architecture, we use full-precision latents, which keeps quantization out of the comparison. In the rate study of Section~\ref{sec:rate}, $b$ is treated as a design variable and sampled during training, so that the same model can be evaluated across several reporting rates.

\subsection{Contention-Based Access and Report Erasure}
\label{sec:uplink}

Reports are sent to the fusion node over shared unreliable reporting links. In many low-cost sensing deployments, receivers access the medium through contention-based or grant-free transmission rather than through dedicated scheduled resources
\cite{abramson1970aloha,shahab2020grantfree}. In this regime, an important impairment is the loss of a whole report. A report may be missing because of a collision, a deep fade, or a missed transmission due to duty-cycle or energy constraints.

We model this impairment as whole-report erasure. Receiver $i$ is delivered to the fusion node with probability $1-\epsilon_i$ and is otherwise dropped:
\begin{equation}
	i\in\mathcal{S}\ \text{with probability}\ 1-\epsilon_i,\qquad
	i\notin\mathcal{S}\ \text{with probability}\ \epsilon_i .
\end{equation}
The erasure pattern is resampled for each observation window. In the experiments, we use a common erasure probability $\epsilon$ unless otherwise stated.

This model preserves the packet-level nature of the reporting link. If a report is erased, its latent never arrives, so the fusion node forms no token for that receiver. The decoder in \eqref{eq:setfusion} must therefore estimate the full emitter set from the surviving reports $\{u_i:i\in\mathcal{S}\}$. This directly connects the access-channel impairment to the set-fusion architecture.

\subsection{Access-Regime Interpretation}
\label{sec:accessmap}

The erasure probability $\epsilon$ summarizes the reliability of the reporting link. It can represent several mechanisms that remove a whole report before it reaches the fusion node, including contention, fading, duty-cycle limits, or energy constraints. As a simple access-level interpretation, consider receivers reporting over a shared grant-free or slotted-ALOHA link with aggregate offered load $G$ report attempts per slot
\cite{shahab2020grantfree}. Under the standard Poisson approximation, a report is delivered if no competing transmission occupies the same slot, giving
\begin{equation}
	\epsilon = 1 - e^{-G}, \qquad T = G e^{-G},
	\label{eq:aloha}
\end{equation}
where $T$ is the aggregate throughput in successful reports per slot. Equation~\eqref{eq:aloha} provides a marginal delivery-probability interpretation of $\epsilon$. The experiments retain independent per-receiver erasures and do not reproduce the correlated collision pattern of a specific random-access protocol. Thus $\epsilon=0.3$ corresponds to $G\approx0.36$, while $\epsilon=0.5$ corresponds to $G\approx0.69$, inside the high-throughput region below the ALOHA peak at $G=1$. This gives a concrete interpretation to the channel-aware training range $\epsilon\in[0,0.5]$. Larger values, such as $\epsilon=0.9$, represent overloaded reporting conditions and serve as stress tests.

\subsection{Channel-Aware Task Objective}
\label{sec:objective}

The encoder and decoder are trained to minimize the expected set distortion over random emitter scenes, receiver geometries, and channel realizations. For a target reporting rate, this can be written as
\begin{equation}
	\min_{\phi,\,\psi}
	\mathbb{E}_{\Theta,\,\mathcal{H}}
	\!\left[D_{\mathrm{set}}(\Theta,\hat{\Theta})\right]
	\quad
	\text{s.t.}\quad
	R_{\mathrm{tot}}\le R_{\mathrm{budget}},
	\label{eq:objective}
\end{equation}
where $\mathcal{H}$ denotes the reporting-link realization. In the erasure experiments, $\mathcal{H}$ is the surviving receiver set $\mathcal{S}$. In the rate study, it also includes the sampled bit depth used for latent quantization.

Placing $\mathcal{H}$ inside the training expectation is what makes the system channel-aware. The model is trained on incomplete report sets, so the encoder learns latents that stay useful when the other reports are missing. The channel-naive baseline occupies the clean-link corner of this objective, with all reports delivered and full-precision latents.

% Section IV: Method  (LOCALIZATION paper)
% Requires: \usepackage{amsmath}

\section{Method}
\label{sec:method}

This section describes the learned reporting architecture, the permutation-invariant set loss, and the channel-aware training procedure. The model has three stages. Each receiver extracts a local time-frequency feature and encodes it into a compact latent. The fusion node aggregates the surviving position-tagged reports as an unordered set. The predicted emitter slots are then matched to the true emitter set through a permutation-invariant distortion.

\subsection{Receiver Feature and Shared Encoder}
\label{sec:encoder}

Receiver $i$ maps its complex baseband window $x_i\in\mathbb{C}^N$ to a time-frequency feature
\begin{equation}
	X_i=\mathcal{T}(x_i).
\end{equation}
In our implementation, $\mathcal{T}$ is a short-time Fourier transform, and the real and imaginary parts are stacked as two input channels. This representation keeps both phase and spectral-envelope information, which are affected by the emitters' nuisance parameters and by propagation.

The feature $X_i$ is passed through a shared convolutional encoder,
\begin{equation}
	z_i=\mathrm{Enc}_{\phi}(X_i)\in\mathbb{R}^{d}.
\end{equation}
The same parameters $\phi$ are used for all receivers. This matters because receiver positions vary from scene to scene, and every receiver plays the same role in the geometry. Sharing the encoder lets the same local feature extractor be applied to any receiver, while the set decoder handles the varying number and membership of reports.

The encoder is implemented as a residual convolutional network followed by global average pooling and a linear projection. Group normalization is used instead of batch normalization, which keeps the encoder independent of batch statistics while erasures are sampled during training. The numerical architecture parameters are reported in the experimental setup.

\subsection{Permutation-Invariant Set-Fusion Decoder}
\label{sec:decoder}

The decoder implements the set map in~\eqref{eq:setfusion}. Each surviving latent is paired with the position of its originating receiver, and the resulting report $u_i=(z_i,r_i)$ is embedded by a shared multilayer perceptron,
\begin{equation}
	h_i=\rho(u_i)=\rho([z_i,r_i]),
	\qquad i\in\mathcal{S},
\end{equation}
where $r_i$ is the normalized receiver position. The embeddings are then aggregated by attention pooling over the surviving set. With a learned query $q$ and keys $k_i$ computed from $h_i$, the pooled summary is
\begin{equation}
	s=\sum_{i\in\mathcal{S}}\alpha_i h_i,
	\qquad
	\alpha_i=
	\frac{\exp(q^\top k_i/\sqrt{d_h})}
	{\sum_{j\in\mathcal{S}}\exp(q^\top k_j/\sqrt{d_h})}.
	\label{eq:attpool}
\end{equation}
Multi-head attention pooling is used in practice, following the Set Transformer pooling block~\cite{lee2019settransformer}.

The aggregation in~\eqref{eq:attpool} is permutation-invariant because the sum and softmax are taken only over the set $\mathcal{S}$. When $\mathcal{S}\neq\emptyset$, erased latents are zeroed and excluded from the attention weights, so an erased report contributes nothing to the summary. For $\mathcal{S}=\emptyset$, the implementation assigns uniform weights to the $M$ inputs with $z_i=\mathbf{0}$. Since the receiver positions remain available, $s$ then depends on receiver geometry alone. This event has probability $\epsilon^{M}$. For $M=4$ it stays rare over most of the training range, at $6.25\%$ for $\epsilon=0.5$ and $1.25\%$ averaged over $\epsilon\sim\mathcal{U}[0,0.5]$, and becomes dominant at the $\epsilon=0.9$ stress point, where it reaches $65.6\%$. The normalized survivor count $|\mathcal{S}|/M$ is appended to $s$, so that the decoder can use the amount of available spatial evidence. A final multilayer perceptron maps the resulting representation to $K_{\max}$ candidate emitter slots,
\begin{equation}
	\hat y_j=(\hat e_j,\hat p_{x,j},\hat p_{y,j}),
	\qquad j=1,\dots,K_{\max},
\end{equation}
where $\hat e_j$ is an existence logit and $(\hat p_{x,j},\hat p_{y,j})$ is the predicted location. Locations are normalized by the region size during training and converted back to meters for reporting.

\subsection{Permutation-Invariant Set Distortion}
\label{sec:loss}

The true emitter set is unordered and has variable cardinality. We therefore match predicted and true slots before computing the loss. Ground truth is padded to $K_{\max}$ slots with existence labels $e_k\in\{0,1\}$. The assignment cost between predicted slot $j$ and target slot $k$ is
\begin{equation}
	C(j,k)=\ell_{\mathrm{BCE}}(\hat e_j,e_k)
	+ e_k\!\!\sum_{c\in\{x,y\}}\!\!\ell_{\mathrm{reg}}(\hat p_{c,j},p_{c,k}),
	\label{eq:assigncost}
\end{equation}
where $\ell_{\mathrm{BCE}}$ is binary cross-entropy and $\ell_{\mathrm{reg}}$ is the smooth-$L_1$ loss. The gate $e_k$ activates the localization term for real emitters, so padded targets carry an existence cost alone.

The optimal assignment is
\begin{equation}
	\pi^\star=\arg\min_{\pi}\sum_{k=1}^{K_{\max}} C(\pi(k),k),
\end{equation}
where $\pi$ ranges over all permutations of the $K_{\max}$ slots. Since $K_{\max}=3$ in the experiments, exhaustive search is sufficient. The training distortion is then
\begin{equation}
	D_{\mathrm{set}}(\Theta,\hat\Theta)
	=\sum_{k=1}^{K_{\max}} C(\pi^\star(k),k).
	\label{eq:setloss}
\end{equation}
The same matching is used to compute the reported metrics: cardinality accuracy and localization RMSE over matched active emitters.

\subsection{Channel-Aware Training and Baselines}
\label{sec:training}

Report erasure is inserted between the receiver encoder and the fusion decoder. For each training minibatch, a surviving set $\mathcal{S}$ is sampled according to the erasure model in Section~\ref{sec:uplink}. The decoder then receives only $\{u_i:i\in\mathcal{S}\}$. This Monte Carlo sampling realizes the expectation over $\mathcal{H}$ in~\eqref{eq:objective} and trains the model under incomplete report sets.

For channel-aware training, the erasure probability is sampled as
\begin{equation}
	\epsilon\sim\mathcal{U}[0,\epsilon_{\max}],
\end{equation}
with $\epsilon_{\max}=0.5$ in the experiments. This trains one model over a range of reporting-link conditions. For the rate study, the bit depth $b$ is also sampled during training, and a straight-through estimator is used through the quantizer. Experiments that isolate erasure and fusion architecture use full-precision latents.

We evaluate three configurations. The channel-aware model uses the set-fusion decoder and is trained with report erasure. The channel-naive model uses the same set-fusion architecture and trains only on clean links, with all reports delivered. The fixed-order baseline shares the receiver encoder and also receives the receiver positions, yet it consumes the $M$ reports in a fixed order in place of set fusion. This baseline is also trained with erasure, and its input dimension stays tied to the receiver count and index ordering used during training. An erased latent becomes a zero vector at the corresponding receiver position, which holds the input dimension fixed.

Both decoders are told which reports arrived: the set decoder through the surviving set $\mathcal{S}$, and the fixed-order baseline through an explicit per-node presence bit appended to its input. The encoder, the available inputs, the training channel, and the reporting budget are therefore held fixed, and the compared models differ in the fusion architecture. The two decoders are not parameter-matched: with $M=4$ and $d=16$, the set-fusion decoder holds $1.9\times10^{5}$ parameters against $1.2\times10^{5}$ for the fixed-order decoder. The relevant evidence is accordingly the way the gap widens with $\epsilon$ in Section~\ref{sec:experiments}, rather than an offset at a single operating point.

% Section V: Experiments and Results  (LOCALIZATION paper)
% Requires: \usepackage{graphicx,booktabs,multirow}
% Figures (in figs/): fig_aware_vs_naive, fig_set_vs_concat, fig_cross_m,
%   fig_rate_reliability  (2 panels each: cardinality acc, localization RMSE)

\section{Experiments and Results}
\label{sec:experiments}

\subsection{Experimental Setup}
\label{sec:setup}

We evaluate the proposed reporting scheme on synthetic multi-emitter
scenes generated according to the model in
Section~\ref{sec:system}. Unless otherwise stated, each scene contains
$M=4$ receivers with positions drawn uniformly at random in an
$1000\times1000$~m region. The cross-count experiment in
Section~\ref{sec:cross_count} uses the same trained model and varies this
number only at evaluation time. Each scene contains $K\in\{1,2,3\}$ active
emitters, drawn uniformly, with random locations and random nuisance
parameters. The occupied bandwidth $B_k$ is drawn uniformly from $1$ to
$16$~MHz, and the center-frequency offset $\nu_k$ is drawn uniformly subject
to the occupied band $[\nu_k-B_k/2,\,\nu_k+B_k/2]$ lying within the
$\pm20$~MHz baseband (sampling rate $40$~MHz); spectral overlap between
emitters is therefore common. The transmit power $P_k$ is drawn uniformly
from $-10$ to $10$~dBm, and the waveform family is either OFDM or
single-carrier. Propagation follows a power-law path loss with exponent
$\eta=2.7$, and all receivers share the same noise power of $-90$~dBm.

Each receiver observes a $1$~ms complex baseband window sampled at $40$~MHz, giving $N=4\times10^4$ samples. The short-time Fourier feature uses a Hann window of length $256$ and hop $160$, which gives $N_t=249$ time frames. The real and imaginary parts are stacked as two channels. The shared encoder produces a $d=16$ dimensional latent for each receiver, and the fusion node estimates the emitter set from the surviving position-tagged reports. Predictions are scored after the set matching of Section~\ref{sec:loss}. We report two task-level metrics: cardinality accuracy, defined as the probability that the predicted count equals $K$, and localization RMSE over the matched active emitters.

\subsection{Signal Generation and Quantizer}
\label{sec:siggen}

The synthetic scenes are generated as follows, so that the dataset can be
reproduced from the description alone.

\emph{Waveforms.} Both families use a QPSK alphabet with unit average symbol
power, drawn uniformly and independently. An OFDM emitter uses symbols of
$L=256$ samples with no cyclic prefix. The number of active subcarriers is
$\mathrm{round}(B_k/f_s\cdot L)$, placed symmetrically about DC with the DC
carrier nulled, and each symbol is formed by an inverse DFT scaled by
$\sqrt{L}$. A single-carrier emitter uses root-raised-cosine pulse shaping with
roll-off $\beta=0.25$ and a span of $8$ symbols, at symbol rate
$R_s=B_k/(1+\beta)$ and $\mathrm{round}(f_s/R_s)$ samples per symbol, with the
filter transient discarded. Each waveform is normalized to unit average power
before transmission.

\emph{Propagation and mixing.} A waveform is generated once per emitter and
shifted to its center-frequency offset by $e^{\jmath 2\pi\nu_k n/f_s}$. It
reaches receiver $i$ with amplitude $\sqrt{P^{\mathrm{lin}}_k}\,\alpha_{i,k}$,
where $P^{\mathrm{lin}}_k=10^{P_k/10}$ converts the transmit power from dBm to
linear units of mW, and
$\alpha_{i,k}=(\max(d_{i,k},d_0)/d_0)^{-\eta/2}$ is the power-law path loss with
$\eta=2.7$ and reference distance $d_0=1$~m. The receiver observation is the sum
over emitters plus circularly-symmetric complex Gaussian noise, drawn
independently per receiver and per sample, whose power $10^{-90/10}$~mW follows
the same linear-unit convention. Only the ratio of these two quantities enters
the problem, so the choice of reference unit leaves the signal-to-noise ratio
unchanged. This model carries
no per-emitter random phase and no propagation delay, so the same complex
waveform reaches every receiver up to a real scale factor. The spatial
information available for localization is therefore the received-power variation
across the $M$ known receiver positions, and time-difference or phase-difference
cues are absent by construction.

\emph{Quantizer.} Each latent component is quantized uniformly to $2^{b}$
levels between per-dimension bounds $z_{\min}$ and $z_{\max}$, by rounding the
normalized value $(z-z_{\min})/(z_{\max}-z_{\min})$ to one of the indices
$\{0,\dots,2^{b}-1\}$. These bounds are
estimated once from clean full-precision latents over $16$ training batches, and
values outside them are clipped. Gradients pass through the quantizer with a
straight-through estimator. In the rate study the bit depth is sampled from
$b\in\{2,3,4,6,8\}$ during training, and the trained model is evaluated at
$b\in\{2,3,4,6,8,16\}$, giving $R_{\mathrm{tot}}=Mbd$ between $128$ and $1024$
bits per window at $M=4$ and $d=16$.

\subsection{Architecture and Training}
\label{sec:archtrain}

The encoder is a residual convolutional network with a strided $5\times5$ stem followed by four residual blocks. The block strides are $(2,2)$, $(2,2)$, $(2,2)$, and $(1,2)$ along the frequency and time axes, with channel widths $(32,32,64,128,128)$. Group normalization is used throughout. The set decoder embeds each surviving report, applies attention pooling, appends the normalized survivor count $|\mathcal{S}|/M$, and uses a final multilayer perceptron with hidden widths $(256,256,128)$ to output $K_{\max}=3$ candidate emitter slots. The dataset contains $40{,}000$ training scenes, $4{,}000$ validation scenes, and $4{,}000$ test scenes.

The evaluation is designed to isolate four factors: channel-aware training, fusion architecture, receiver-count generalization, and semantic rate. The channel-aware and channel-naive models use the same set-fusion architecture. The fixed-order baseline uses the same receiver encoder and receives receiver positions for fairness, but replaces the set decoder with a fixed-order concatenation decoder. All models are trained for $50$ epochs with AdamW, learning rate $10^{-3}$, weight decay $10^{-4}$, batch size $64$, and gradient-norm clipping at $1.0$. The learning rate follows a cosine schedule with a five-epoch linear warmup and is annealed to zero by the final epoch. The checkpoint with the lowest validation loss is used for testing. At evaluation, each operating point is averaged over $30$ stochastic channel realizations with fixed seeds.

The setup is deliberately scoped: the number of emitters is limited to $K\le3$, the waveform set contains two families, the receiver noise power is identical across nodes, and the propagation model is synthetic. These choices keep the comparison focused on the reporting channel and the fusion rule. More realistic propagation, denser scenes, and geometry-dependent receiver reliability are left for future work.

\subsection{Channel-Aware Co-Design}

We first evaluate the effect of exposing the semantic encoder and fusion decoder to report erasures during training. The channel-aware and channel-naive models use the same shared encoder and permutation-invariant decoder; they differ only in the training channel. The former samples erasures during training, with $\epsilon \sim \mathcal{U}[0,0.5]$, whereas the latter is trained only on clean links. Both models are then evaluated under increasing erasure probability $\epsilon$.

\begin{figure}[t]
  \centering
  \includegraphics[width=\columnwidth]{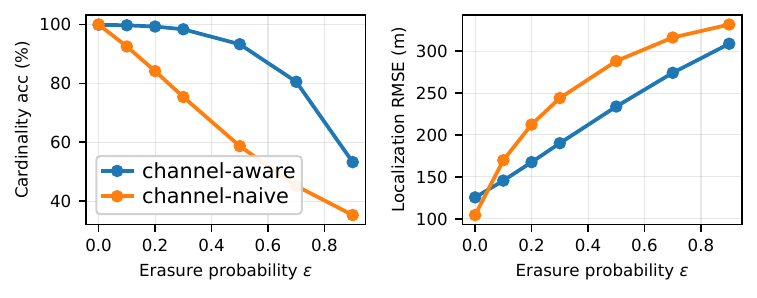}
 \caption{Channel-aware versus channel-naive set-fusion models under report erasure: cardinality accuracy (left) and localization RMSE (right). The channel-naive model performs well on clean links but degrades rapidly as reports are erased. Channel-aware training preserves counting accuracy over a wider erasure range and gives lower localization error once loss is appreciable.}
 \label{fig:aware_vs_naive}
\end{figure}

Fig.~\ref{fig:aware_vs_naive} shows that channel-aware training is essential once report loss becomes appreciable. On clean links, both models perform well, and the channel-naive model holds a slight localization advantage because it is optimized for the full-report regime. As $\epsilon$ increases, the channel-naive model degrades rapidly. Its cardinality accuracy drops sharply, which reveals a decoder that relies on the joint availability of several receiver reports. The channel-aware model instead maintains high counting accuracy over a much wider erasure range and degrades gracefully at high loss.

The same trend appears in localization. The channel-naive model has the lowest RMSE only in the near-clean regime, but its localization error increases quickly as reports are erased. The channel-aware model crosses over at low-to-moderate erasure and remains consistently better thereafter. This confirms that placing the erasure channel inside the training expectation reshapes the learned representation: each report becomes useful on its own, rather than only in the simultaneous presence of the other receivers. The gain therefore appears as robustness in the unreliable reporting-link regime considered in this paper, which is where it matters.

\subsection{Set Fusion versus Fixed-Order Concatenation}

We next isolate the effect of the fusion rule. The proposed decoder treats the surviving reports as an unordered set, whereas the fixed-order baseline concatenates the receiver latents in a prescribed order. Both models use the same shared receiver encoder, both receive the receiver positions and an explicit indication of which reports arrived, and both are trained with channel-aware erasure under the same reporting budget. The encoder, the available inputs, the training channel, and the budget are therefore held fixed, and the models differ in how the fusion node represents the collection of reports.

\begin{figure}[t]
  \centering
  \includegraphics[width=\columnwidth]{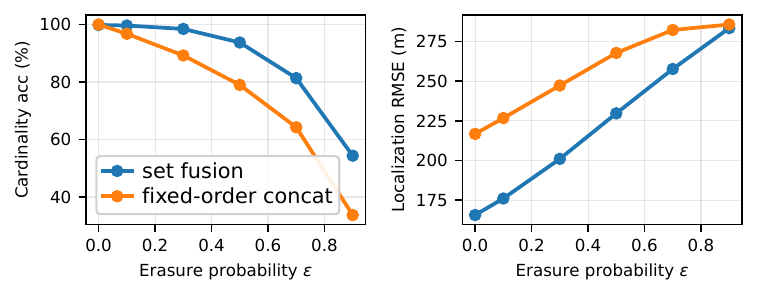}
\caption{Permutation-invariant set fusion versus fixed-order concatenation under report erasure. Both models are trained with channel-aware erasure and both receive receiver-position information. Set fusion is more robust for cardinality estimation and gives lower localization RMSE across the erasure sweep. The observed advantage is consistent with treating erased reports as absent set elements rather than as missing entries in a fixed-order representation.}
  \label{fig:set_vs_fixed}
\end{figure}

Fig.~\ref{fig:set_vs_fixed} shows that the permutation-invariant decoder is substantially more robust to report loss. On clean links, both models achieve high counting accuracy, but the set decoder already gives lower localization error. As erasure increases, the difference becomes more pronounced for cardinality estimation: the fixed-order decoder degrades rapidly, while the set decoder preserves high counting accuracy over a much wider range of erasure probabilities. This behavior is consistent with the structure of the problem. In the set decoder, an erased report is simply absent from the input set, and the pooling operation is performed over the reports that actually arrive. The fixed-order decoder instead retains a fixed-dimensional, ordered input, so missing reports perturb a representation built for constant membership.

For localization, set fusion also remains better across the sweep. The gap is largest in the clean and moderate-erasure regimes, where the decoder can exploit the spatial diversity of several surviving receivers. At very high erasure, the two localization curves approach each other because both models are limited by the small number of reports that remain. The main conclusion is therefore twofold: set fusion improves robustness under report loss, and, more importantly, it matches the reporting-link structure by making the decoder operate on the surviving receiver set in place of a fixed concatenation.

\subsection{Deployment Across Receiver Counts}
\label{sec:cross_count}
We now test whether the proposed set-fusion model can operate beyond the receiver count used during training. The channel-aware model is trained only with $M=4$ deployed receivers and is then evaluated, without retraining, at $M \in \{2,3,4,6\}$. This sweep suits the permutation-invariant formulation: a fixed-order concatenation decoder fixes its input dimension at the training count and index ordering, so it accepts $M=4$ alone.

\begin{figure}[t]
  \centering
  \includegraphics[width=\columnwidth]{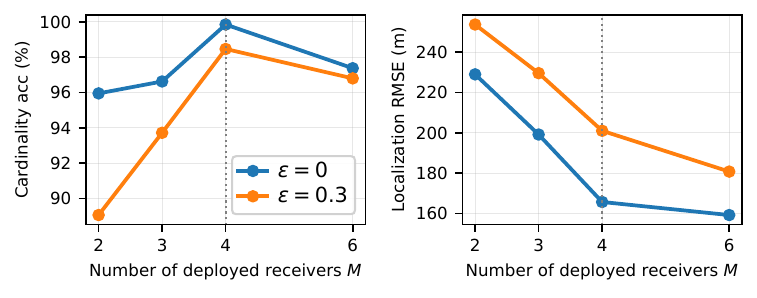}
\caption{Cross-count deployment of one channel-aware set-fusion model trained at $M=4$ deployed receivers (dotted line) and evaluated without retraining at $M \in \{2,3,4,6\}$. Localization improves as more receivers are deployed, both on clean links and under erasure. Cardinality accuracy remains high but is best near the training configuration. A fixed-order concatenation decoder cannot be evaluated here because its input dimension is tied to the training receiver count.}
  \label{fig:cross_count}
\end{figure}

Fig.~\ref{fig:cross_count} shows that the same model remains effective across receiver counts. Localization improves as more receivers are deployed, both on clean links and under erasure. This trend is expected: additional receivers provide more spatial diversity and make the inverse localization problem better constrained. The improvement is especially clear when moving from two to four receivers. Increasing the count beyond the training value still improves localization, although with diminishing returns.

Cardinality estimation also remains strong across receiver counts, and peaks near the training configuration. The drop at $M=2$ reflects the limited spatial evidence available from a few receivers, especially when erasure is present. The mild drop at $M=6$ shows that extrapolating beyond the training count remains possible at a small cost. The distinction matters: performance tracks the receiver count rather than staying identical across it, while one architecture and one checkpoint cover the whole range.

The practical implication is that the proposed decoder supports deployment flexibility. Receivers can be added, removed, or temporarily lost, and the fusion node still processes the surviving set of reports. Performance then follows the amount of spatial information actually available, in place of a fixed input format.

\subsection{Rate--Reliability}
\label{sec:rate}

We finally study how the semantic reporting rate interacts with report loss. The previous experiments used full-precision latents in order to isolate the effects of erasure and fusion architecture. Here, each latent component is quantized to $b$ bits, and the total reporting load is $R_{\mathrm{tot}}=Mbd$ bits per observation window. With $M=4$ and $d=16$, this corresponds to $64b$ bits across all receivers.

\begin{figure}[t]
  \centering
  \includegraphics[width=\columnwidth]{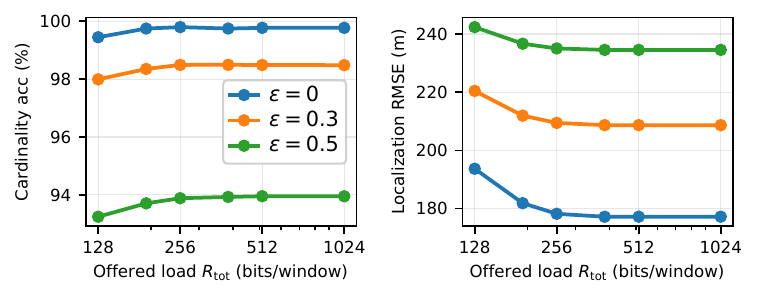}
\caption{Rate--reliability behavior of the semantic reporting scheme. Cardinality accuracy (left) and localization RMSE (right) are shown versus total reporting load $R_{\mathrm{tot}}=Mbd$ for erasure probabilities $\epsilon \in \{0,0.3,0.5\}$. Performance saturates after a few bits per latent component, around $R_{\mathrm{tot}}\approx 256$ bits for $M=4$ and $d=16$. Erasure shifts the achievable performance level but does not substantially move the rate knee in the tested range.}

  \label{fig:rate_reliability}
\end{figure}

Fig.~\ref{fig:rate_reliability} shows cardinality accuracy and localization RMSE as a function of the total offered reporting load $R_{\mathrm{tot}}$ for several erasure probabilities. The main observation is that only a few bits per latent component are needed. Performance improves rapidly at low rates and then saturates around a few hundred bits per window, with little additional gain beyond approximately $b=4$ bits per component, i.e., $R_{\mathrm{tot}}\approx 256$ bits for four receivers.

Erasure mainly changes the performance level while the rate at which saturation occurs holds. Higher erasure probabilities reduce the achievable cardinality accuracy and increase the localization error, as expected, because fewer receiver reports reach the fusion node. The saturation knee nevertheless remains in roughly the same rate range across the tested erasure levels. This suggests that, in this operating regime, quantization and report loss act as partly separable impairments: increasing the bit depth beyond the knee leaves the missing spatial observations unrecovered.

The resulting reporting load is small compared with raw forwarding. At $b=4$, each receiver sends a 64-bit semantic report, or 256 bits in total for $M=4$. Forwarding the raw complex window would require $N \times 32 \approx 1.28$~Mbit per receiver with 16-bit I/Q samples, or $5.12$~Mbit in total across four receivers. Comparing the two aggregates, the semantic representation reduces the reporting payload by roughly $2\times 10^4$. Fig.~\ref{fig:rate_reliability} shows little additional task benefit from raising the latent precision beyond this point, which is the operating claim made here; the comparison quantifies payload, and it does not establish equivalence with a centralized raw-I/Q receiver.

\subsection{Discussion}
\label{sec:discussion}

The results indicate that semantic reporting is useful because it addresses two constraints at once: limited reporting capacity and unreliable links. A few hundred task-relevant bits per observation window suffice in the tested setting, a payload several orders of magnitude below forwarding raw I/Q samples. Report erasure also has to enter the design a priori. A channel-naive model degrades quickly when reports are lost, while channel-aware training makes the learned reports more useful under partial observation.

The fusion rule is equally important. Since the link determines which reports arrive, the decoder should operate on the surviving set in place of a fixed-order vector. This explains both the robustness of set fusion under erasure and its ability to operate at receiver counts different from the one used during training. The rate results further show that quantization and erasure play different roles: quantization limits the information carried by each surviving report, while erasure limits the amount of spatial evidence available to the fusion node. Beyond the rate knee, increasing the bit depth leaves the missing receiver observations unrecovered.

\section{Conclusion}
\label{sec:conclusion}

We presented a semantic communication scheme for distributed spectrum monitoring over unreliable links. Receivers send compact latent reports, and a permutation-invariant decoder pairs each survivor with its receiver position to estimate the emitter count and locations from whatever crosses the channel. Training with report erasure inside the task objective improves robustness, while set fusion outperforms fixed-order concatenation and supports deployment across receiver counts. The rate study shows that four bits per latent component suffice in the setting considered here, or $256$ bits per observation window at $M=4$.

The present evaluation covers synthetic propagation, at most three emitters, two waveform families, identical receiver noise, and independent per-node erasures. Future work will extend it to denser scenes, more realistic propagation, correlated and geometry-dependent erasures, and secondary estimation of spectral nuisance parameters.

\bibliographystyle{IEEEtran}
\bibliography{refs}

@article{gunduz2023beyond,
  author  = {G{\"u}nd{\"u}z, Deniz and Qin, Zhijin and Aguerri, I{\~n}aki Estella and Dhillon, Harpreet S. and Yang, Zhaohui and Yener, Aylin and Wong, Kai Kit and Chae, Chan-Byoung},
  title   = {Beyond Transmitting Bits: Context, Semantics, and Task-Oriented Communications},
  journal = {IEEE J. Sel. Areas Commun.},
  volume  = {41},
  number  = {1},
  pages   = {5--41},
  year    = {2023}
}

@article{strinati2021beyond,
  author  = {Calvanese Strinati, Emilio and Barbarossa, Sergio},
  title   = {{6G} Networks: Beyond {S}hannon Towards Semantic and Goal-Oriented Communications},
  journal = {Comput. Netw.},
  volume  = {190},
  pages   = {107930},
  year    = {2021}
}

@article{shao2022ib,
  author  = {Shao, Jiawei and Mao, Yuyi and Zhang, Jun},
  title   = {Learning Task-Oriented Communication for Edge Inference: An Information Bottleneck Approach},
  journal = {IEEE J. Sel. Areas Commun.},
  volume  = {40},
  number  = {1},
  pages   = {197--211},
  year    = {2022}
}

@article{shao2022dib,
  author  = {Shao, Jiawei and Mao, Yuyi and Zhang, Jun},
  title   = {Task-Oriented Communication for Multi-Device Cooperative Edge Inference},
  journal = {IEEE Trans. Wireless Commun.},
  volume  = {22},
  number  = {1},
  pages   = {73--87},
  year    = {2023}
}

@article{bourtsoulatze2019djscc,
  author  = {Bourtsoulatze, Eirina and Kurka, David Burth and G{\"u}nd{\"u}z, Deniz},
  title   = {Deep Joint Source-Channel Coding for Wireless Image Transmission},
  journal = {IEEE Trans. Cogn. Commun. Netw.},
  volume  = {5},
  number  = {3},
  pages   = {567--579},
  year    = {2019}
}

@article{dai2022ntscc,
  author  = {Dai, Jincheng and Wang, Sixian and Tan, Kailin and Si, Zhongwei and Qin, Xiaoqi and Niu, Kai and Zhang, Ping},
  title   = {Nonlinear Transform Source-Channel Coding for Semantic Communications},
  journal = {IEEE J. Sel. Areas Commun.},
  volume  = {40},
  number  = {8},
  pages   = {2300--2316},
  year    = {2022}
}

@article{xie2021dl,
  author  = {Xie, Huiqiang and Qin, Zhijin and Li, Geoffrey Ye and Juang, Biing-Hwang},
  title   = {Deep Learning Enabled Semantic Communication Systems},
  journal = {IEEE Trans. Signal Process.},
  volume  = {69},
  pages   = {2663--2675},
  year    = {2021}
}

@inproceedings{stavrou2022rd,
  author    = {Stavrou, Photios A. and Kountouris, Marios},
  title     = {A Rate Distortion Approach to Goal-Oriented Communication},
  booktitle = {Proc. IEEE Int. Symp. Inf. Theory (ISIT)},
  pages     = {590--595},
  year      = {2022}
}

@article{tishby2000ib,
  author  = {Tishby, Naftali and Pereira, Fernando C. and Bialek, William},
  title   = {The Information Bottleneck Method},
  journal = {arXiv preprint physics/0004057},
  year    = {2000}
}

@inproceedings{zhan2021deepmtl,
  author    = {Zhan, Caitao and Ghaderibaneh, Mohammad and Sahu, Pranjal and Gupta, Himanshu},
  title     = {{DeepMTL}: Deep Learning Based Multiple Transmitter Localization},
  booktitle = {Proc. IEEE Int. Symp. World of Wireless, Mobile and Multimedia Networks (WoWMoM)},
  year      = {2021}
}

@article{rajendran2018dl,
  author  = {Rajendran, Sreeraj and Meert, Wannes and Giustiniano, Domenico and Lenders, Vincent and Pollin, Sofie},
  title   = {Deep Learning Models for Wireless Signal Classification With Distributed Low-Cost Spectrum Sensors},
  journal = {IEEE Trans. Cogn. Commun. Netw.},
  volume  = {4},
  number  = {3},
  pages   = {433--445},
  year    = {2018}
}

@misc{distcomp2026,
  author        = {Bicer, H. Nazim and Laneman, J. Nicholas},
  title         = {Spatially Distributed Task-Oriented Compression for Multi-Emitter Localization and Characterization with Spectral Overlap},
  howpublished  = {arXiv:2606.01446 [eess.SP]},
  year          = {2026},
  eprint        = {2606.01446},
  archivePrefix = {arXiv},
  primaryClass  = {eess.SP}
}

@inproceedings{abramson1970aloha,
  author    = {Abramson, Norman},
  title     = {The {ALOHA} System---Another Alternative for Computer Communications},
  booktitle = {Proc. Fall Joint Computer Conf. (AFIPS)},
  pages     = {281--285},
  year      = {1970}
}

@article{shahab2020grantfree,
  author  = {Shahab, Muhammad Basit and Abbas, Rana and Shirvanimoghaddam, Mahyar and Johnson, Sarah J.},
  title   = {Grant-Free Non-Orthogonal Multiple Access for {IoT}: A Survey},
  journal = {IEEE Commun. Surveys Tuts.},
  volume  = {22},
  number  = {3},
  pages   = {1805--1838},
  year    = {2020}
}

@article{akyildiz2011coop,
  author  = {Akyildiz, Ian F. and Lo, Brandon F. and Balakrishnan, Ravikumar},
  title   = {Cooperative Spectrum Sensing in Cognitive Radio Networks: A Survey},
  journal = {Physical Communication},
  volume  = {4},
  number  = {1},
  pages   = {40--62},
  year    = {2011}
}

@inproceedings{sun2007coop,
  author    = {Sun, Chunhua and Zhang, Wei and Letaief, Khaled Ben},
  title     = {Cooperative Spectrum Sensing for Cognitive Radios under Bandwidth Constraints},
  booktitle = {Proc. IEEE Wireless Commun. Netw. Conf. (WCNC)},
  pages     = {1--5},
  year      = {2007}
}

@inproceedings{tian2007compressed,
  author    = {Tian, Zhi and Giannakis, Georgios B.},
  title     = {Compressed Sensing for Wideband Cognitive Radios},
  booktitle = {Proc. IEEE Int. Conf. Acoust., Speech, Signal Process. (ICASSP)},
  pages     = {1357--1360},
  year      = {2007}
}

@inproceedings{madden2002tag,
  author    = {Madden, Samuel and Franklin, Michael J. and Hellerstein, Joseph M. and Hong, Wei},
  title     = {{TAG}: A {T}iny {AG}gregation Service for Ad-Hoc Sensor Networks},
  booktitle = {Proc. USENIX Symp. Operating Systems Design and Implementation (OSDI)},
  pages     = {131--146},
  year      = {2002}
}

@inproceedings{zaheer2017deepsets,
  author    = {Zaheer, Manzil and Kottur, Satwik and Ravanbakhsh, Siamak and P{\'o}czos, Barnab{\'a}s and Salakhutdinov, Ruslan and Smola, Alexander J.},
  title     = {Deep Sets},
  booktitle = {Advances in Neural Information Processing Systems (NeurIPS)},
  pages     = {3391--3401},
  year      = {2017}
}

@inproceedings{lee2019settransformer,
  author    = {Lee, Juho and Lee, Yoonho and Kim, Jungtaek and Kosiorek, Adam R. and Choi, Seungjin and Teh, Yee Whye},
  title     = {Set Transformer: A Framework for Attention-Based Permutation-Invariant Neural Networks},
  booktitle = {Proc. Int. Conf. Machine Learning (ICML)},
  pages     = {3744--3753},
  year      = {2019}
}

\end{document}